\documentclass[aps,pra,twocolumn,showpacs,groupedaddress]{revtex4}

\usepackage{graphicx}
\usepackage{amsmath}

\newcommand*{\ket}[1]{\mathop{\left| #1 \right\rangle}\nolimits}

\begin{document}

\title{Interrogation of caesium atoms in a fountain clock\\ by a femtosecond laser microwave oscillator}

\author{S.~Weyers, B.~Lipphardt, and H.~Schnatz}

\affiliation{Physikalisch-Technische Bundesanstalt, Bundesallee 100, 38116 Braunschweig, 
Germany}

\date{\today}

\begin{abstract}
A caesium fountain clock is operated utilizing a microwave oscillator that 
derives its frequency stability from a stable laser by means of a fiber-laser 
femtosecond frequency comb. This oscillator is based on the technology 
developed for optical clocks and replaces the quartz based microwave 
oscillator commonly used in fountain clocks. As a result, a significant 
decrease of the frequency instability of the fountain clock is obtained, 
reaching $0.74 \times 10^{-14}$ at $100\,$s averaging time. We could 
demonstrate that for a significant range of detected atom numbers the 
instability is limited by quantum projection noise only, and that for 
the current status of this fountain clock the new microwave source poses no 
limit on the achievable frequency instability.
\end{abstract}

\pacs{06.30.Ft, 42.62.Eh, 42.50.Lc}

\maketitle

\noindent Progress in important fields of fundamental physics entails the request for better and better clocks \cite{Peik}. The advent of optical clocks necessitates frequency measurements against caesium fountain clocks, which nowadays deliver the best results for the realization of the SI second in terms of systematic and statistical uncertainty \cite{Wyn}. In fountain clocks laser cooled atoms are vertically tossed, exposed to two microwave pulses during the ballistic flight and finally detected in a state selective detection process. The microwave frequency is adjusted to the atomic transition frequency according to the detection signal. Relative systematic uncertainties below the $10^{-15}$-level have already been achieved by several fountains \cite{Wyn}. In the best case, frequency instabilities are limited by the quantum projection noise \cite{Itano, Santarelli2}, resulting in $10^{-16}$ statistical uncertainties after one day of measurement \cite{Vian}. 
However, for the majority of fountain clocks the frequency instability is eventually limited by the phase noise properties of the employed quartz oscillators through the Dick effect \cite{Dick}. This effect results from an aliasing of interrogation oscillator frequency noise components caused by the non-continuous probing of the atomic transition frequency in commonly pulsed fountains (see e.g. \cite{Santarelli1}). With the best  available quartz oscillators as ``flywheels'' and for typical fountain duty cycles the relative frequency instability is thus limited to the order of $10^{-13} (\tau/$s$)^{-1/2}$, where $\tau$ is the measurement time. A possible, but complex and demanding way to essentially avoid a frequency stability degradation due to this effect is the switch to a fountain operating with a continuous beam of cold atoms \cite{Joyet}. As an elaborate  solution for the established pulsed fountains a highly stable cryogenic sapphire oscillator \cite{Luiten} has been used as interrogation oscillator, resulting in a fountain frequency instability in the low $10^{-14} (\tau/$s$)^{-1/2}$ range \cite{Vian}. 

Another approach is the employment of an optically generated microwave \cite{McF,Kim}, for which we use as a flywheel a laser originally set up for use as interrogation laser in the Yb$^+$ frequency standard of the Physikalisch-Technische Bundesanstalt (PTB) \cite{Schnei}. In the experiment reported here, this laser is locked to a stable cavity but not to the Yb$^+$ ion. This reference laser reaches a short-term frequency instability of a few times $10^{-15}$ in $1\,$s, with a long-term drift of $10^{-16}$/s due to the optical cavity. This instability is transferred to a dielectric resonator oscillator (DRO) at a frequency $\nu_\mathrm{\,DRO}=9.6\,$GHz which is our highly stable microwave source for the caesium fountain \cite{Lipphardt}. A self-referenced fiber-laser femtosecond frequency comb stabilized to a hydrogen maser connects the optical and microwave spectral regions. The DRO is phase-locked to the reference laser using the frequency comb as a transfer oscillator \cite{Lipphardt}. The independently measured frequency instability of the locked DRO is $10^{-14}$ in $1\,$s, and reaches the level of the reference laser after $10\,$s. 

\begin{figure}
\includegraphics[width=8.6cm]{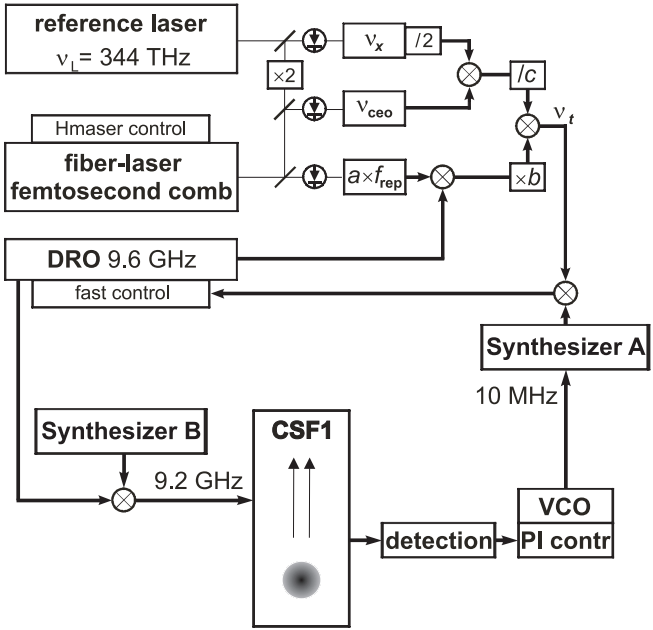}
\caption{Scheme of the setup showing in the upper part the fast servo loop for the microwave oscillator utilizing an ultrastable laser and a fiber-laser femtosecond frequency comb (FLFC). In the lower part the slow closed-loop control of the DRO (via the synthesizer A) by the caesium fountain signal is shown. $\nu_x$: beat note of the $m$th mode of the FLFC and the reference laser with the mode number $m=a\times b\times c$ ($a,b\,$: integer numbers; $c\,$: rational number); $\nu_\mathrm{ceo}$: offset frequency of the FLFC; $f_\mathrm{rep}$: repetition rate of the FLFC.
\label{fig:SetupScheme}}
\end{figure}

Figure~\ref{fig:SetupScheme} is a schematic of the DRO lock and the long-term stabilization of the microwave frequency to the caesium clock transition. The transfer beat frequency $\nu_t$ results from a direct phase comparison between the DRO and the reference laser \cite{Lipphardt}:

\begin{equation}
\nu_t = \nu_L/c-b\; \nu_\mathrm{\,DRO},
\label{eq:FLMO}
\end{equation}

\noindent
where $\nu_L = 344\,$THz is the Yb$^+$ reference laser frequency, $c$ a rational division and $b$ an integer  multiplication factor (see  Fig.~\ref{fig:SetupScheme}). A phase-locked loop technique steers the DRO to keep $\nu_t$ constant at the frequency given by synthesizer A. Synthesizer B and a single-sideband mixer shift the DRO frequency to the interrogation frequency for the fountain. A detection and integration process gives the correction signal for the microwave frequency. A slow proportional-integral (PI) controller drives synthesizer A with a $10\,$MHz voltage controlled oscillator (VCO) via the reference frequency input. This compensates the drift of the reference laser and defines the mean frequency of the DRO. We have utilized this  femtosecond laser microwave oscillator (FLMO) instead of the usual quartz based microwave oscillator (QBMO) for the interrogation of atoms in the PTB caesium fountain clock CSF1 \cite{Weyers1}. 

If interrogation oscillator noise contributions can be neglected, the frequency instability of a fountain expressed by the Allan standard deviation is given as (see e.g. \cite{Santarelli2}):

\begin{equation}
\sigma_y(\tau)=\frac{1}{\pi Q_\mathrm{at}\; S/N}\;\sqrt{\frac{T_c}{\tau}},
\label{eq:sigma}
\end{equation}

\noindent 
where $\tau$ is the measurement time, $T_c$ the fountain cycle duration, $Q_\mathrm{at}=\nu_0/\Delta \nu$ is the atomic quality factor with $\Delta \nu$ the width of the Ramsey resonance and $\nu_0$ the caesium hyperfine frequency. $S/N$ is the signal-to-noise ratio, which is proportional to the square-root of the number of detected atoms $N_\mathrm{at}=N_3+N_4$ in case of quantum projection noise contributing only. $N_4$ and $N_3$ represent the number of detected atoms in the caesium ground state hyperfine levels $\ket{F=4,m_F=0}$ and $\ket{F=3,m_F=0}$, respectively.

\begin{figure}
\includegraphics[width=8.6cm]{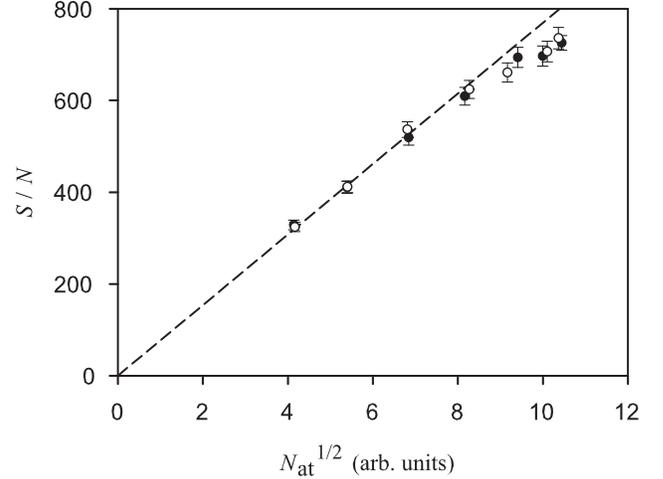}
\caption{Signal-to-noise ratio $S/N$ as a function of the square-root of the detected atom number $N_\mathrm{at}$ utilizing the FLMO. $\circ\,$: $2\times\pi/4$-pulses at $\nu_0$; $\bullet\,$: $2\times\pi/2$-pulses at $\nu_0+\Delta \nu/2$. From the dashed line, which indicates the quantum projection noise limit, we derive that $N_\mathrm{at}=1$ (in arbitrary units) corresponds to a number of $\approx 6000$ detected atoms. 
\label{fig:SignaltoNoise}}
\end{figure}

We first use the free running FLMO for applying two $\pi/4$-pulses in resonance at a frequency $\nu=\nu_0$ to the atoms in CSF1, when the slow closed-loop control of the DRO is suspended (see  Fig.~\ref{fig:SetupScheme}). In this case the influence of interrogation oscillator noise on the detected atomic population is effectively suppressed. For each fountain cycle the normalized clock signal $S=N_4/(N_3+N_4)$ is obtained. $S$ corresponds to the amplitude of the central Ramsey fringe and is independent of $N_\mathrm{at}$ due to the normalization. By changing the loading time of the magneto-optical trap used as source of atoms in CSF1, the number of cooled and launched atoms, and thus $N_\mathrm{at}$, can be varied. For a range of loading times between $0.16\,$s and $3.16\,$s ($T_c=1.12\,\mathrm{s}\mbox{--} 4.12\,$s), $S$ is recorded during 1000 fountain cycles. Subsequently the Allan standard deviations $\sigma_S(N_\mathrm{at})$ of the single $S$ measurements are calculated. For the different loading times we obtain ratios $S/N=S/\sigma_S(N_\mathrm{at})$, which are shown as a function of the square-root of $N_\mathrm{at}$ by the open circles in Fig.~\ref{fig:SignaltoNoise}. As indicated by the dashed line the $S/N$-ratios are essentially proportional to the square-root of $N_\mathrm{at}$ as it is expected for quantum projection noise contributing only. This confirms that for the given useful range of $N_\mathrm{at}$ noise contributions due to the photon shot-noise and the detection system can be neglected in CSF1 \cite{Weyers3}. 

In order to check for noise contributions of the FLMO, we compare the latter kind of operation with an operation mode which entails the same quantum projection noise contribution, but is also sensitive to interrogation oscillator noise. Using the free-running FLMO for applying two $\pi/2$-pulses to the atoms at a frequency detuning $\Delta \nu/2$ from resonance results in the same measured signal $S$ as in the preceding measurements. Actually this kind of operation corresponds to the normal servo-controlled operation of CSF1 as a clock where the central Ramsey fringe is alternately probed at frequency detunings $\pm\Delta \nu/2$ and both, quantum projection and interrogation oscillator noise, contribute. The resulting $S/N$-ratios are depicted by the solid black circles in Fig.~\ref{fig:SignaltoNoise}. The fact that both data sets (open and solid black circles) coincide well indicates that for the given useful range of $N_\mathrm{at}$ the FLMO does not add any significant noise in the clock operation mode.

\begin{figure}
\includegraphics[width=8.6cm]{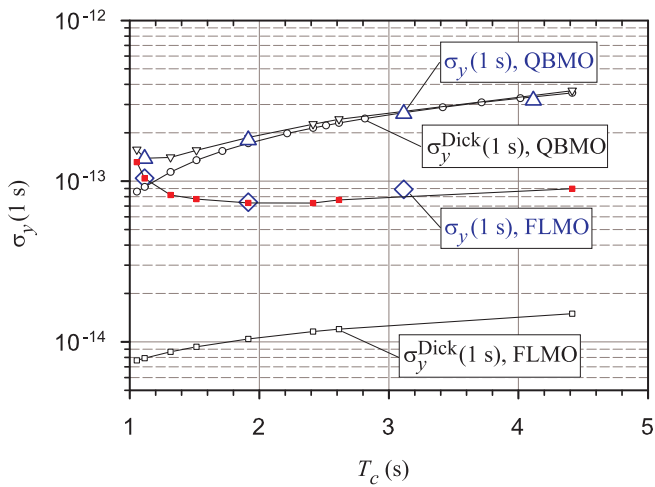}
\caption{(Color online) Calculated (small symbols) and measured (large symbols) Allan standard deviations $\sigma_y(1\,\mathrm{s})$ of relative frequency differences $y$(CSF1-Hmaser) as a function of the cycle time $T_c$ for different noise contributions. Solid squares: calculated data in the ideal case of vanishing  oscillator noise contributions; triangle down symbols: calculated data in the case of additional noise contributions due to the QBMO. The solid lines are for guiding the eye. See text for further explanations. 
\label{fig:SigmaAllan}}
\end{figure}

We now discuss the findings in the case when the slow closed-loop control of the DRO is active. The solid squares in Fig.~\ref{fig:SigmaAllan} indicate the expected frequency instability $\sigma_y(1\,\mathrm{s})$ of CSF1 calculated in the case of negligible interrogation oscillator noise contributions. For this calculation measured $S/N$-ratios for two resonant $\pi/4$-pulses were used as input for Eq.~\eqref{eq:sigma}. In order to obtain $S/N$-ratios for a range of $N_\mathrm{at}$, the loading time of the magneto-optical trap, and thus $T_c$ was varied. An increasing loading time involves an exponential saturation to a maximum of $N_\mathrm{at}$ with a time constant of $\approx 1\,$s. The decrease of $\sigma_y(1\,\mathrm{s})$ for cycle times up to $2.4\,$s results from the dominating effect of the increased $S/N$-ratio due to the increased  atom number $N_\mathrm{at}$ accompanying extended loading times [see Eq.~\eqref{eq:sigma}]. The increase of $\sigma_y(1\,\mathrm{s})$ for cycle times longer than $2.4\,$s is due to the now dominating contribution of the square root of the cycle time $T_c$ in Eq.~\eqref{eq:sigma}, as the necessary loading time for increasing $N_\mathrm{at}$ furthermore increases disproportionately.

When CSF1 operates as a frequency standard the pulsed mode of operation entails a significant amount of dead time, mainly caused by the time needed for loading and detecting of the atoms, during which the interrogation oscillator frequency is not controlled by the atomic resonance signal. This dead time gives rise to a degradation of the fountain frequency instability caused by the frequency noise of the interrogation oscillator (Dick effect). The corresponding contribution to the Allan standard deviation is given by \cite{Santarelli1}:

\begin{equation}
\sigma_{y}^\mathrm{Dick}(\tau)=\left(\frac{1}{\tau}\sum_{m=1}^{\infty}\frac{g_m^2}{g_0^2}\;S_y^f(m/T_c)\right)^{1/2},
\label{eq:sigmaoscillator}
\end{equation}

\noindent 
where $S_y^f(m/T_c)$ is the one-sided power spectral density of the relative frequency fluctuations of the free running interrogation oscillator at Fourier frequencies $m/T_c$, and $g_m$ are the Fourier coefficients of the Fourier series expansion of the sensitivity function $g(t)$ \cite{Santarelli1}, which is the response of the atomic system to a phase step of the interrogation oscillator at time $t$.

The open circles in Fig.~\ref{fig:SigmaAllan} show the calculated frequency instability contributions due to the Dick effect according to Eq.~\eqref{eq:sigmaoscillator}, when the QBMO is in use for CSF1. The calculation is based on the phase noise data from the test sheet of the actual quartz crystal oscillator employed 
\cite{Syf} and the sensitivity function $g(t)$ calculated for CSF1. With increasing $T_c$ an increasing number of oscillator phase noise components at small Fourier frequencies $m/T_c$ degrade the frequency stability. The triangle down symbols show the quadratic sum of both instability contributions calculated according to  Eq.~\eqref{eq:sigma} and Eq.~\eqref{eq:sigmaoscillator}. A comparison of this calculated $\sigma_y(1\,\mathrm{s})$ with frequency instabilities of CSF1 (large triangle up symbols) obtained from frequency measurements with the QBMO using a hydrogen maser as a reference exhibit a good agreement. In Fig.~\ref{fig:SigmaAllan} and in the following all frequency instability data given by measurements is obtained by subtracting small frequency instability contributions of the reference maser by using the three-corner-hat method \cite{3chat}, taking into account the results of simultaneous frequency measurements against a second hydrogen maser.

If we now consider the calculated $\sigma_y^\mathrm{Dick}(1\,\mathrm{s})$ due to the Dick effect, when the FLMO is in use for CSF1, we notice that the resulting frequency instability contributions shown by the open squares in Fig.~\ref{fig:SigmaAllan} are at or even below $10^{-14}$. For this calculation we use the measured phase noise of the FLMO \cite{Lipphardt} and again $g(t)$, as calculated for CSF1. The results suggest that the frequency instability contribution due to the FLMO is negligible compared to $\sigma_y(1\,\mathrm{s})$ calculated for the case of quantum projection noise contributing only (solid squares in Fig.~\ref{fig:SigmaAllan}). 

This is confirmed by measurements of the frequency of CSF1 against a hydrogen maser, using the FLMO for the interrogation of the atoms. The resulting measured frequency instabilities of CSF1 alone are depicted in Fig.~\ref{fig:SigmaAllan} as large open diamonds. The agreement of this experimental data with the expected frequency instability (solid squares in Fig.~\ref{fig:SigmaAllan}) is conclusive. Using the FLMO for the atom interrogation in CSF1 yields a purely quantum projection noise limited frequency instability down to $0.74 \times 10^{-13} (\tau/$s$)^{-1/2}$. The uninterrupted frequency measurements used for this evaluation typically lasted one day. Continuous measurements of several days demonstrated the reliability of the complete system. 

\begin{figure}
\includegraphics[width=8.6cm]{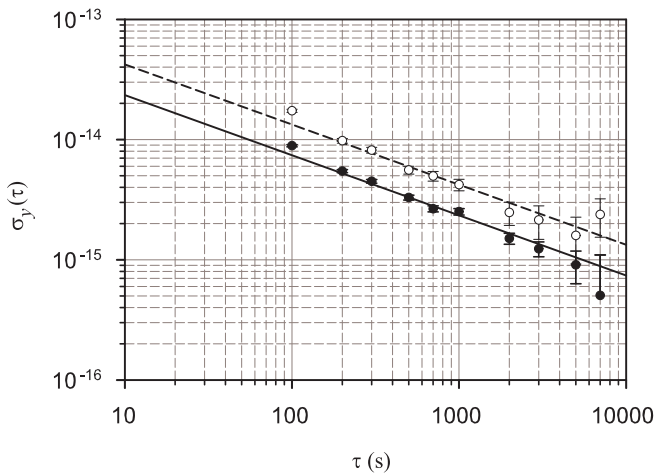}
\caption{Measured Allan standard deviations $\sigma_y(\tau)$ of the frequency of CSF1 using the QBMO~($\circ$) or the FLMO~($\bullet$) for interrogating the atoms. The solid and the dashed line show frequency instabilities $\sigma_y(\tau)=1.33 \times 10^{-13} (\tau/$s$)^{-1/2}$ and $\sigma_y(\tau)=0.74 \times 10^{-13} (\tau/$s$)^{-1/2}$, respectively.
\label{fig:SigmaAllanEx}}
\end{figure}

The improvement in fountain clock stability is also demonstrated by comparing the measured frequency instabilities $\sigma_y(\tau)$ of CSF1 depicted in Fig.~\ref{fig:SigmaAllanEx}. The open data points are obtained using the QBMO and choosing the loading time $T_{c}\approx 1.1\,$s in order to achieve the minimum frequency instability (see triangle down symbols in Fig.~\ref{fig:SigmaAllan}). The dashed line in Fig.~\ref{fig:SigmaAllanEx} indicates an instability of $1.33 \times 10^{-13} (\tau/$s$)^{-1/2}$. The solid black data points are obtained with the FLMO and a correspondingly optimized loading time, resulting in  $T_{c}\approx 1.9\,$s. The frequency instability of $0.74 \times 10^{-13} (\tau/$s$)^{-1/2}$ indicated by the solid line shows that the measurement time needed to obtain a specific statistical uncertainty of CSF1 is shortened by a factor of 3.2 compared to the case using the QBMO. Moreover, in the case of collisional shift measurements which involve the need for operating CSF1 at elevated $N_\mathrm{at}$ \cite{Weyers1} the gain in measurement time can be even larger. In order to further lower the frequency instability of CSF1, refined methods for loading more atoms in a given time interval, e.g., by loading from a cold beam of atoms, need to be installed. 

In summary, employing an optically generated microwave in a caesium fountain clock resulted in a reduced frequency instability of the clock below the $10^{-14}$-level at $100\,$s averaging time only limited by quantum projection noise. This approach avoids the cost of operation of a low-noise cryogenic sapphire oscillator and is thus an alternative solution for reducing the frequency instability of fountain clocks. A low instability is indispensable for the evaluation of several systematic uncertainty contributions at the level of $10^{-16}$ or below. Moreover, the statistical uncertainty of frequency measurements of optical clocks using fountain clocks as references is limited by fountain frequency instabilities and will highly benefit from quantum projection noise limited fountain instabilities. In fact, the obtained instability of $0.74 \times 10^{-13} (\tau/$s$)^{-1/2}$ enables us to reach a statistical uncertainty of $2.5\times 10^{-16}$ after one day of measurement.

We thank Chr.~Tamm for making available the ultrastable ytterbium reference laser, gratefully acknowledge useful discussions with R.~Schr\"oder, and the critical reading of the manuscript by R.~Wynands and E.~Peik.

\pagebreak


\begin{thebibliography}{99}
\bibitem{Syf} The one-sided power spectral density of the relative frequency fluctuations is $S_y^f(f) = 1.6\times 10^{-26}/f + 1.6\times 10^{-27}f/\mathrm{Hz}^2 + 1.0\times 10^{-29}f^2/\mathrm{Hz}^3$.
\end{thebibliography}
\end{document}